\documentclass[pra,reprint]{revtex4-1}
\usepackage{amsmath,amssymb,graphicx}

\usepackage{xcolor}

\renewcommand{\vec}{\mathbf}

\newcommand{\const}{\mathop{\rm const}\nolimits}

\graphicspath{{source/}}

\begin{document}

\title{Optical beam diffraction  tensor in birefringent crystals}

\author{Konstantin B. Yushkov}\email{Corresponding author: konstantin.yushkov@misis.ru}
\author{Natalya F. Naumenko}
\affiliation{National University of Science and Technology ``MISIS'', 4 Leninsky Prospekt, Moscow 119049, Russia}

\begin{abstract}
We demonstrate that anisotropy of Fresnel diffraction in a birefringent medium can be quantitatively characterized by a planar tensor. Eigenvectors of the tensor correspond to directions of minimum and maximum beam divergence. Zero eigenvalues exist for the fast eigenmode near the optic axis of biaxial crystals and correspond to autocollimated beam propagation. Applications of the beam diffraction tensor are demonstrated for analysis of noncritical phase matching in acousto-optics and three-wave mixing.
\end{abstract}

%\pacs{42.60.Fc, 42.65.Re, 42.79.Jq}

\maketitle %% required

\section{Introduction}

Birefringence is the fundamental phenomenon of crystal optics~\cite{YarivYeh}. Its applications span through various areas of optical engineering  and emerging fields of photonics. Optical anisotropy of crystals is widely used in polarization optics, acousto-optics, laser physics and nonlinear optics. Periodic structures made of anisotropic materials can exibit properties of phoxonic (i.e. simultaneously photonic and phononic) crystals~\cite{RollandDupontGazalet14}. Conical refraction in biaxial crystals enables generation of Bessel beams~\cite{PetitJolySegondsBoulanger13,TurpinLoikoKalkandjievMompart16} and specific geometries of acousto-optic diffraction can be used for controlling them~\cite{BelyiKhiloKazak16}.

The wave normal surface describes the directional dependence of the material refractive index $n$ (real part of the dielectric permittivity tensor $\widehat\varepsilon$) in $k$-space. In gyrotropic crystals, the shape of the normal surface is also affected by the gyration tensor. The normal surface is used to describe beam propagation, diffraction, and interaction phenomena. Other anisotropic optical properties of crystals are absorption (associated with the imaginary part $\widehat\varepsilon$) and fluorescence~\cite{PetitJolySegondsBoulanger13}.

Fundamental results in the field of birefringent optics include generalization of the Huygens principle to anisotropic media~\cite{BergsteinZachos66,Ogg71}. Based on this principle and vector diffraction theory, paraxial Gaussian beam propagation in uniaxial crystals has been described analytically~\cite{FleckFeit83,CiattoniCrosignaniPorto01,Nawareg19}. The result of related work can be briefly summarized as follows: slowly varying amplitudes of the two eigenwaves satisfy uncoupled parabolic equations and interference of those components at arbitrary input polarization may result in sufficient transformations of the beam profile.

The most obvious effect of birefringence is energy walk-off, i.e. noncollinearity between the wave vector and the Poynting vector. The Poynting vector is parallel to the group velocity, that is to the surface normal vector. Thus, the magnitude of walk-off is related to the first derivative of the normal surface. Unless the normal surface (or one of its sheets) is a sphere, its second derivatives are not identically zero. As it will be demonstrate, those derivatives determine beam diffraction in anisotropic media.

The beam diffraction tensor was introduced in crystal acoustics by Khatkevich~\cite{Khatkevich78_eng} and Naumenko \emph{et al.}~\cite{NaumenkoPerelomovaBondarenko83_eng}. It also helps to find singularities of the wave normal surface (conical axes), which is a problem in acoustics since in crystal systems with trigonal or lower symmetry acoustic axes may exist in arbitrary directions not restricted to crystal symmetry planes~\cite{NaumenkoYushkovMolchanov21}. Analysis of the wave normal surface geometry by Alshits and Lyubimov showed that the same approach can be applied both for optical and acoustic waves in crystals~\cite{AlshitsLyubimov13_eng}.

In this work, we develop the concept of the beam diffraction tensor to crystal optics and demonstrate its applications to analysis of noncritical phase matching (NPM) in acousto-optics and nonlinear optics. It is shown that geometrical analysis of the slowness surface provides a quick path for estimation of beam propagation parameters both in uniaxial and biaxial crystals and quantitatively describes coupling-of-modes in non-paraxial cases. Relevant applications include wide-angle configurations of acousto-optic tunable filters (AOTFs), second harmonic generation (SHG), third harmonic generation (THG), and optical parametric amplification (OPA).

\section{Beam diffraction tensor}

\subsection{Definition and calculation}
The beam diffraction tensor $\widehat W$ is defined as~\cite{NaumenkoYushkovMolchanov21}
\begin{equation}\label{eq-W}
 W_{ij} = \delta_{ij} - s_i s_j + g_i g_j + \frac{\partial g_i}{\partial s_j}
\end{equation}
where $\delta_{ij}$ is the Kronecker delta, $\vec s$ is the wave normal unity vector, and $\vec g$ is the normalized transverse group velocity vector. The group velocity vector is
\begin{equation}\label{eq-Vg}
  \vec v_{\rm g} = \frac{d v}{d \vec s}.
\end{equation}
where $v=c/n(\vec s)$ is the phase velocity in the wave normal direction $\vec s$. Then, $g$ is defined as
\begin{equation}\label{eq-g}
  \vec g = \vec v_{\rm g}/v - \vec s.
\end{equation}

For numerical calculations it is more convenient to use the property that $\vec v_{\rm g}$ is collinear to the Poynting vector $\vec S$ since it helps to replace numerical differentiation~\eqref{eq-Vg} with explicit calculation. The alternative and equivalent expression for the  group velocity vector is
\begin{equation}\label{eq-Vg2}
  \vec v_{\rm g} = \frac{v\vec S}{|\vec S|\cos\beta}
\end{equation}
where $\beta$ is the walk-off angle.

Further procedure for calculating $\vec S$ is based on the method of index ellipsoid~\cite[Ch.~4]{YarivYeh}. The impermeability tensor $\widehat\eta = \widehat\varepsilon^{-1}$ is transformed to the frame of reference with $\vec s$ as one of the coordinate axes. This makes possible to reduce the eigenvalue problem from three dimensions to two. The electrical field induction $\vec D$ is an eigenvector of transverse impermeability tensor $\widehat\eta_{\rm t}$. At this step, we calculate refractive indices and polarizations of both eigenwaves in the crystal, select one of them, and repeat the following steps for both of them separately. Then, electrical field can be found as contraction of the full impermeability tensor with the induction vector $\vec D$:
\begin{equation}\label{eq-E}
  \vec E = \widehat \eta \vec D.
\end{equation}
In crystal optics, $(\vec D \vec s)=0$ and $(\vec E \vec S)=0$ that allows calculating $\beta$ as the angle between $\vec D$ and $\vec E$, or
\begin{equation}\label{eq-gamma}
  \cos\beta = \frac{(\vec D \vec E)}{|\vec D||\vec E|}.
\end{equation}
In turn, the Poynting vector can be represented as
\begin{equation}\label{eq-S}
  \vec S = |\vec E|^2 \vec s - (\vec E\vec s) \vec E.
\end{equation}
 A constant scaling factor is omitted since only direction of $\vec S$ is effective  in~\eqref{eq-Vg2}.

Equations \eqref{eq-Vg}--\eqref{eq-S} give the path for analytical calculation of $g(\vec s)$. Then, $\widehat W(\vec s)$ is calculated from (\ref{eq-W}) using finite differencing.

\subsection{Eigenvalues}

\begin{figure}
  \centering
  \includegraphics[width=\columnwidth]{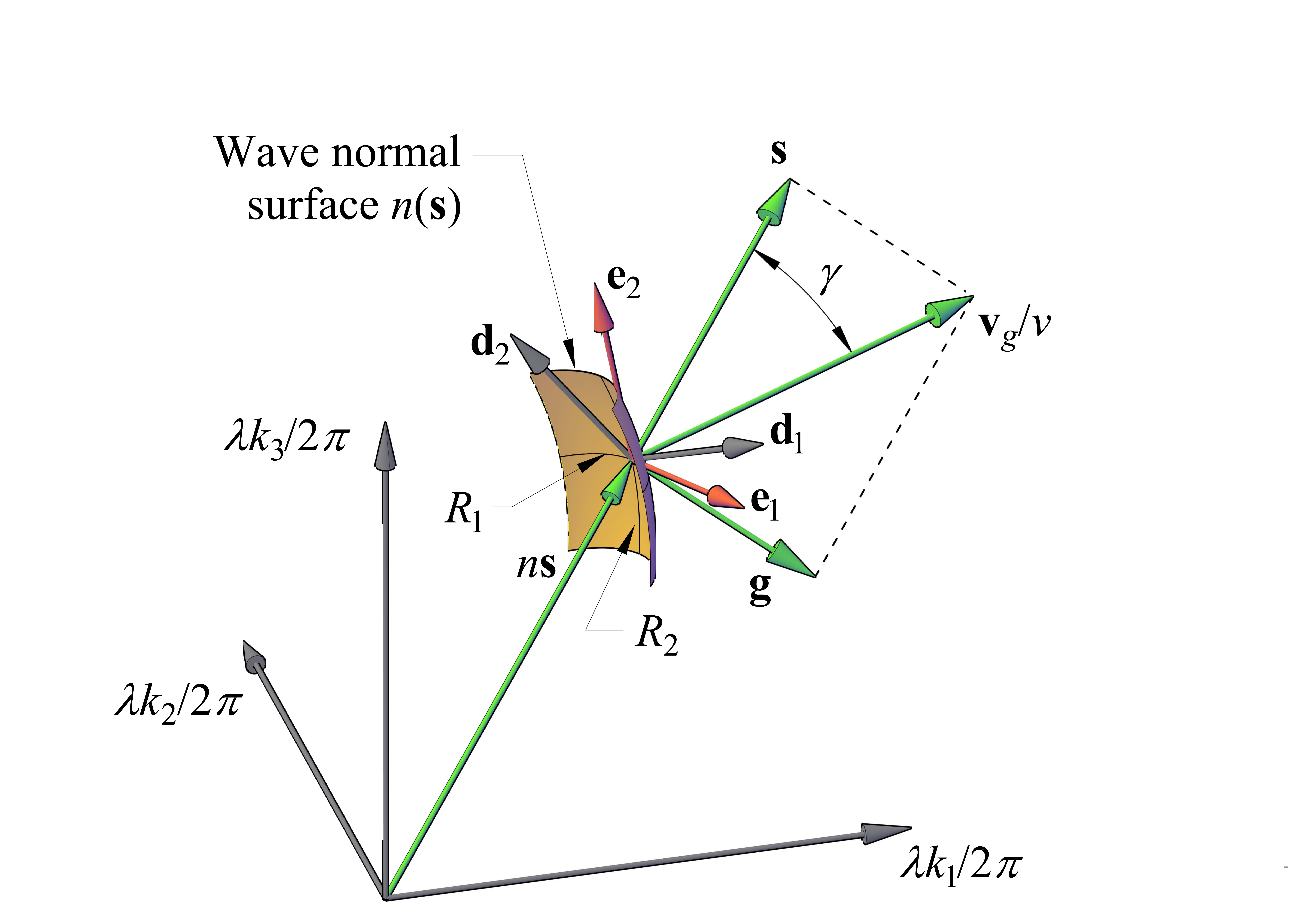}
  \caption{Eigenvalues and eigenvectors of the diffraction tensor $\widehat W$ correspond to principal curvatures of the wave normal surface.}\label{fig-sch}
\end{figure}

The diffraction tensor $\widehat W$ is a planar second rank tensor, i.e. $\vec s \widehat W \vec s = 0$. It can be diagonalized and has two real eigenvalues $w_i$ ($i=1,2$). For the reason discussed in Sec.~\ref{sec-Frau}, we will call them \emph{diffraction coefficients}. Two eigenvectors of $\widehat W$, $\vec d_1$ and $\vec d_2$, are orthogonal to $\vec s$, the third eigenvector is $\vec d_3=\vec s$ with 0 eigenvalue. There are two eigenvalues for each wave normal direction $\vec s$ and for each optical eigenmode. The local geometry of the normal surface is schematically shown in Fig.~\ref{fig-sch}. The vectors $\vec e_i$ ($i=1,2$) are the principal directions and belong to the tangential plane, i.e. $(\vec e_i \vec v_{\rm g})=0$.

The diffraction coefficients $w_i(\vec s)$ have a simple geometrical interpretation. They are proportional to principal curvatures of the wave normal surface with $n(\vec s)$ as the proportionality factor. As it is known from differential geometry, the principal curvatures are the maximum and the minimum curvatures of any regular surface and they belong to orthogonal planes~\cite{Carmo}. The principal directions $\vec e_i$ together with the surface unit normal vector $\vec v_{\rm g}/|\vec v_{\rm g}|$ form an orthonormal basis,  which can be obtained by a transformation of the orthonormal basis formed by $\vec d_i$ and the direction vector $\vec s$. In some applications, it can be convenient to use wave normal surface curvature radii
\begin{equation}\label{eq-ri}
  R_i(\vec s) = \frac{n(\vec s)}{w_i(\vec s)}
\end{equation}
instead of diffraction coefficients.

The eigenvalues of $\widehat W$ can be either positive, negative, or zero. In the following section, it is proved that the eigenvalues are related to divergence of the optical beam in Fresnel approximation. Particularly, $w_i=0$ corresponds to diffraction-free propagation of the beam (strictly, the beam does not diverge only in the direction of the corresponding eigenvector $\vec d_i$), the phenomenon known as \emph{autocollimation}.

The  optic axes in biaxial crystals are conical degeneracies, i.e. the directions where the wave normal surface is not differentiable. One on the eigenvalues for each mode is unbound when approaching the direction of the optic axis, and the corresponding curvature radius vanishes. On the contrary, the optical axis in uniaxial crystals belongs to a topologically different type of degeneracy (tangential), and the slowness surface remains smooth. Topology of the wave normal surface in the vicinity of the optic axes and singularities of the polarization field are the same as in crystal acoustics~\cite{AlshitsLyubimov13_eng,Shuvalov98}.

%\subsection{Relation between $\widehat W$ and paraxial diffraction pattern}~\label{sec-Frau}

\section{Fresnel diffraction in anisotropic medium}~\label{sec-Frau}

The eigenvalues of  $\widehat W$ are normalized so that their absolute values are proportional to the far field beam divergence. The proof of this fact is the same as for acoustic waves since it relies on general principles of vector field diffraction in parabolic approximation~\cite{IUS13}. Hereinafter, we briefly reproduce this proof.

In scalar diffraction theory, the electromagnetic field $A(\vec r)$ can be expressed as a convolution integral~\cite{BornWolf7ed,GoodmanIFO}
\begin{equation}\label{eq-a1}
   A(\vec r) = \int  A_0 (\vec r') G(\vec r - \vec r') d\vec r'
\end{equation}
where $A_0$ is the source field, and $G$ is the Green's function. Explicit equation for $G(\vec r)$ depends on the properties of the medium and the assumptions made.

We assume that the source is planar and located at $z=0$, the radiation is monochromatic with the vacuum wavelength $\lambda$ and the wavenumber $k=2\pi n/\lambda$, and the wave vector $\vec k_0 = k\vec s$ corresponds to $z$ direction. The Green's function of free space in Fresnel approximation is
 \begin{equation}\label{eq-G-free}
   G(\vec r) = \frac{i k \exp(i k z)}{2\pi z}\exp\left[i\frac{k(x^2+y^2)}{2z}\right].
 \end{equation}

To find the Green's function in anisotropic medium we recall representation of $A(\vec r)$ through the two-dimensional Fourier transform
\begin{equation}\label{eq-a4}
   A(\vec r) = \int \widetilde A (\vec k) \exp[i(\vec k \vec r)] d^2 \vec k
\end{equation}
where
\begin{equation}\label{eq-a5}
    \widetilde A (\vec k) = \frac{1}{(2\pi)^2}\int A_0 (\vec r')\exp[-i(\vec k \vec r')] d^2\vec r'
\end{equation}
is the angular spectrum at the source plane $z=0$. Then,
\begin{equation}\label{eq-a6}
   G(\vec r) = \frac{1}{(2\pi)^2} \int \exp[i(\vec k \vec r) ] d^2\vec k.
\end{equation}
In paraxial approximation, the wave vector can be represented as $\vec k = \vec k_0 + \vec q$, and $|\vec q|\ll|\vec k_0|$.
The scalar product $(\vec k \vec r)$ in \eqref{eq-a6} can be expanded in the Taylor series in two dimensions, $q_x$ and $q_y$:
\begin{equation}\label{eq-a7}
  (\vec k \vec r) = k z + q_i(r_i - g_i z) - \frac{1}{2k}W_{ij}q_iq_j z + \dots
\end{equation}
where summation over repeated indices of transverse components $i,j=x,y$ is assumed,
\begin{equation}\label{eq-a8}
  g_i  = - \left.\frac{\partial k_z}{\partial q_i}\right|_{q_i=0}
\end{equation}
and
\begin{equation}\label{eq-a9}
  W_{ij}  = - k \left.\frac{\partial^2 k_z}{\partial q_i \partial q_j}\right|_{q_i,q_j=0}
\end{equation}
Finally, substituting~\eqref{eq-a7} into~\eqref{eq-a6} we obtain
\begin{multline}\label{eq-a10}
   G(\vec r) = \frac{1}{(2\pi)^2}\exp(i k z)\times \\
   \iint \exp[iq_i(r_i - g_i z)] \exp\left[-\frac{i z}{2 k} W_{ij}q_iq_j \right]d q_x d q_y.
\end{multline}
The variables are separable and~\eqref{eq-a10} can be easily integrated if $W_{ij}=0$ for $i\neq j$, i.e. in the frame of reference where the tensor $\widehat W$ is diagonal. The result is
\begin{multline}\label{eq-a11}
   G(\vec r) = \frac{i k}{2\pi z \sqrt{W_{xx} W_{yy}}}\exp(i k z)\times\\
   \exp\left\{i\frac{k}{2z}\left[\frac{(x-g_x z)^2}{W_{xx}} + \frac{(y-g_y z)^2}{W_{yy}}\right]\right\}.
\end{multline}

Comparing~\eqref{eq-G-free} and~\eqref{eq-a11} one can see that $g_x$ and $g_y$ describe linear shift of the beam along transverse directions $x$ and $y$ respectively. Thus, linear series coefficients $g_i$ are responsible for the beam walk-off, and from the definition~\eqref{eq-a8} one can derive that they are equal to components of $\vec g$, hence:
\begin{equation}\label{eq-9}
  \sqrt{g_x^2 + g_y^2} = |\vec g| = \tan\beta
\end{equation}
The quadratic coefficients $W_{xx}$ and $W_{yy}$ describe scaling of the Green's function $A$ along $x$ and $y$ directions. Substituting $k_z=k(\vec s) s_z$ and $q_i = k(\vec s) s_i$ into~\eqref{eq-a9}, one can easily verify with the help of~\eqref{eq-g} and~\eqref{eq-Vg} that the definitions~\eqref{eq-W} and~\eqref{eq-a9} are equivalent. Besides that, we have chosen $x$ and $y$ coordinates so that the tensor $\widehat W$ is diagonal, therefore $W_{xx}=w_1$ and $W_{yy}=w_2$.

\begin{figure}
  \centering
  \includegraphics[width=\columnwidth]{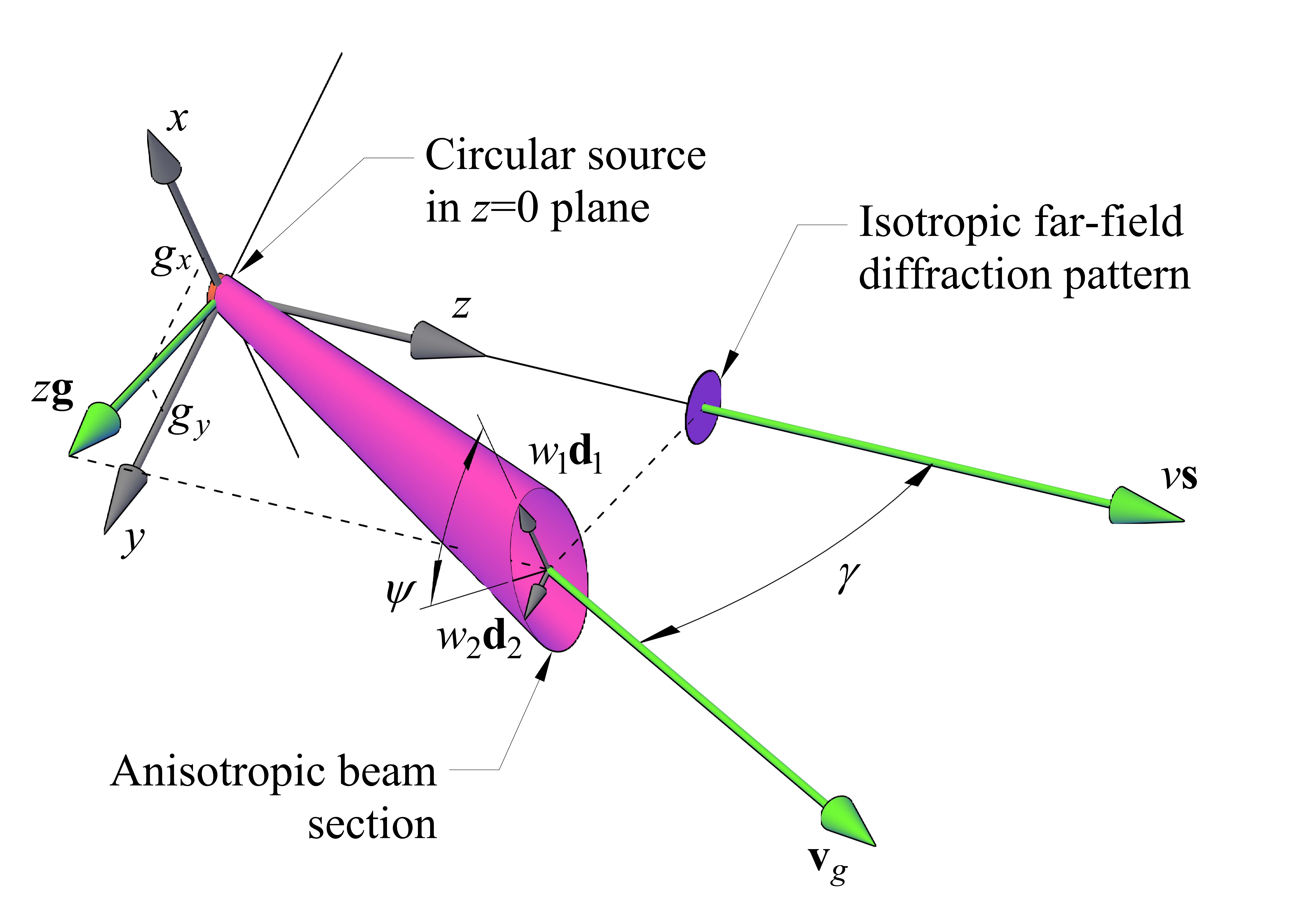}
  \caption{Transformation of far-field diffraction pattern in anisotropic media.}\label{fig-trans}
\end{figure}

In isotropic media $g_x=g_y=0$ and $W_{xx}=W_{yy}=1$. The transformation of the beam with symmetrical source according to the Green's function \eqref{eq-a11} is schematically shown in Fig.~\ref{fig-trans}.

Calculation of the convolution integral~\eqref{eq-a1} for a Gaussian beam demonstrates that the sign of $W_{ii}$ affects only the phase of $A(\vec r)$. Thus, one can introduce the diffraction length for a Gaussian beam with a waist radius $\rho_0$ as
\begin{equation}\label{eq-Ld}
  L_i = \frac{k \rho_0^2}{|W_{ii}|}.
\end{equation}
The radius of the beam is different in directions of $x$ and $y$ and can be found as
\begin{equation}\label{eq-rho}
  \rho_i(z) = \rho_0\sqrt{1 + (z/L_i)^2}.
\end{equation}
In the far field, $\rho_i(z)\approx z |W_{ii}|/(k \rho_0)$, i.e. it is $|W_{ii}|$ times greater than in the isotropic medium.

\section{Applications}

\subsection{Autocollimation and effect on beam structure}\label{sec-auto}

Explicit expression \eqref{eq-a11} for the Green's function $ G(\vec r)$ was obtained under assumption that $W_{xx} W_{yy}\neq 0$. This assumption however does not necessarily hold, and one of the coefficients can be zero. Without loss of  generality, we can take $w_{2}=W_{yy}=0$ and $w_{1}=W_{xx}\neq 0$.  In this case, we use integral expression for the Dirac delta function $\delta(y)$ and obtain
\begin{multline}\label{eq-G-auto}
   G(\vec r) = \sqrt{\frac{i k}{2\pi z w_{1}}}\exp(i k z)\times\\
   \exp\left[i\frac{k}{2z}\frac{(x-g_x z)^2}{w_{1}}\right] \delta(y - g_y z).
\end{multline}
This means that there is no beam profile transformation along $y$ axis.

\begin{figure}[t]
  \centering
  \includegraphics[width=\columnwidth]{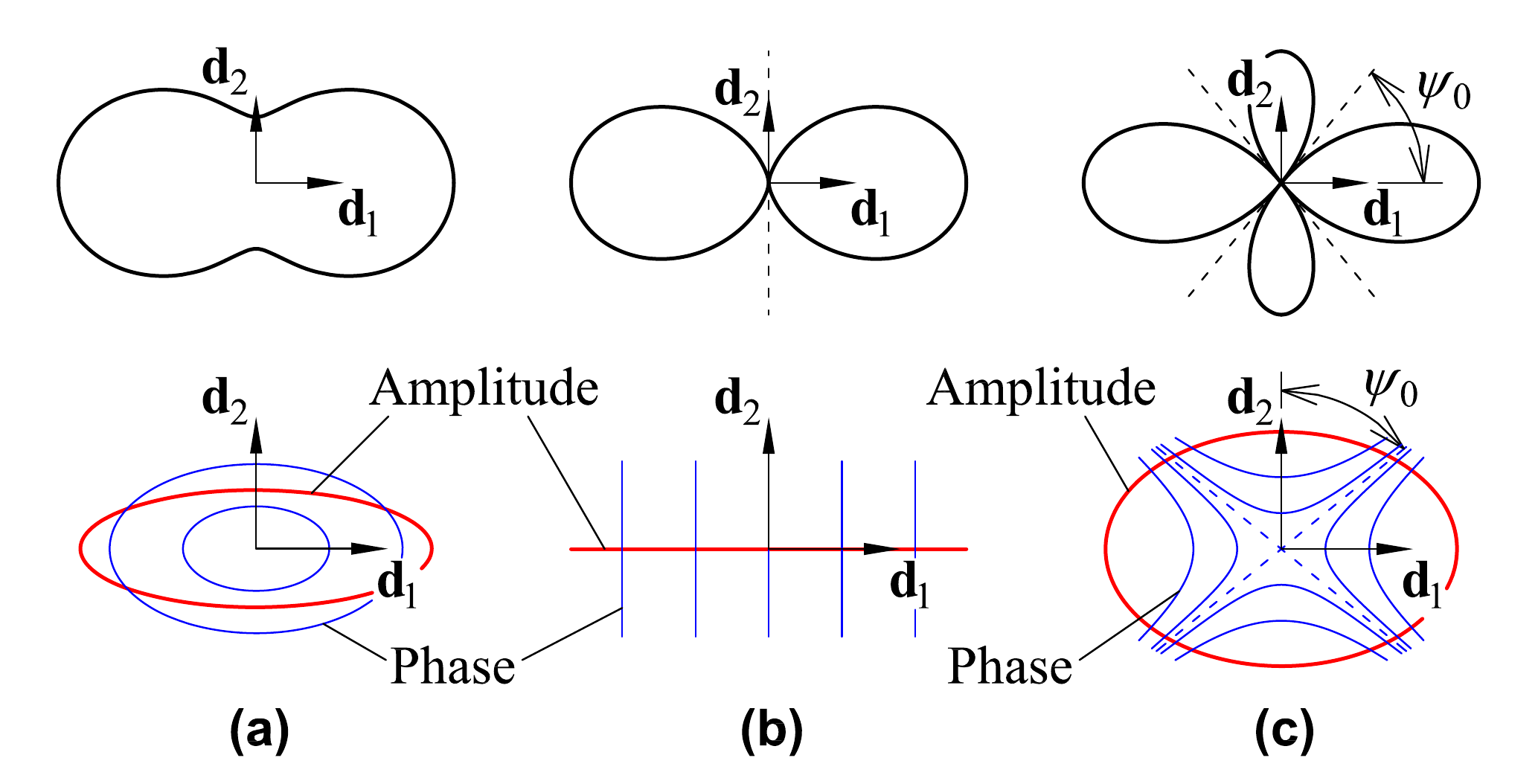}
  \caption{Spatial pattern of $w(\psi)$ values (top) and far-field equal phase and amplitude contours (bottom): (a) $w_1 w_2 >0$; (b) $w_1 w_2 =0$; (c) $w_1 w_2 < 0$.}\label{fig-Wpsi}
\end{figure}

According to tensor transformation rules $w(\psi)$ can be obtained for an arbitrary transverse direction defined by local azimuth angle $\psi$ (see Fig.~\ref{fig-trans}):
\begin{equation}\label{eq-Wpsi}
  w(\psi) = w_1 \cos^2 \psi + w_2 \sin^2 \psi
\end{equation}
The directions $\psi_0$ corresponding to $w(\psi_0)=0$ can be found as
\begin{equation}\label{eq-psi0}
  \psi_0 = \pm \arctan\sqrt{-\frac{w_1}{w_2}}.
\end{equation}
Typical spatial patterns of $w(\psi)$ at different signs of $w_1$ and $w_2$ are shown in Fig.~\ref{fig-Wpsi}.

The effect of diffraction tensor on beam structure can be easily demonstrated for a Gaussian beam. In the near field, $z\ll L_i$, the phase of the beam is expressed as
\begin{equation}\label{eq-Phi-near}
\Phi(x,y) = \frac{z}{2 k \rho_0}\left[w_{1}(x - g_x z)^2 + w_{2}(y - g_y z)^2\right].
\end{equation}
In the far field, $z\gg L_i$, the phase is
\begin{equation}\label{eq-Phi-far}
\Phi(x,y) = \frac{k}{2 z}\left[\frac{(x - g_x z)^2}{w_{1}} +\frac{(y - g_y z)^2}{w_{2}}\right].
\end{equation}
In both cases, the phase front $\Phi(x,y) =\const$ is an elliptic paraboloid when $w_1 w_2>0$ and a hyperbolic paraboloid when $w_1 w_2<0$. In optics, the case when both $w_1$ and $w_2$ are negative does not exist, and Fig.~\ref{fig-Wpsi}(a) corresponds to a convex wavefront. The cases of autocollimation, Fig.~\ref{fig-trans}(b), and different signs of $w_i$, Fig.~\ref{fig-trans}(c), exist only for the slow eigenmode in biaxial crystals. In the latter case, the saddle point of the wavefront is $x=g_x z$ and $y=g_y z$, but the directions of phase isolines are different for the near field and for the far field.

In the near field, the amplitude distribution $|A(\vec r)|$ does not change, and the phase is modulated so that phase isolines satisfy the equation
\begin{equation}\label{eq-near-Phase}
  w_1 x^2 + w_2 y^2 =\const.
\end{equation}
Thus, the phase does not change along the lines $w(\psi)=0$.

In the far field, the amplitude isolines are the ellipses the principal axes ratio $|w_x/w_y|$:
\begin{equation}
  \frac{x^2}{w_1^2} + \frac{y^2}{w_2^2} = \const
\end{equation}
Those isolines do not depend on the sign of $w_x$ and $w_y$. The phase isolines are the
\begin{equation}
  \frac{x^2}{w_1} + \frac{y^2}{w_2} = \const
\end{equation}
They are ellipses with the principal axes ratio of $\sqrt{w_x/w_y}$ if $w_x w_y>0$ and hyperbolas if $w_x w_y<0$ as shown in Fig.~\ref{fig-Wpsi}. Note that zero phase isolines now make the angle of $\pi/2-\psi_0$ with $\vec{d}_1$ axis and does not correspond with the directions of $w(\psi)=0$.

\begin{figure}[t]
  \centering
  \includegraphics[width=\columnwidth]{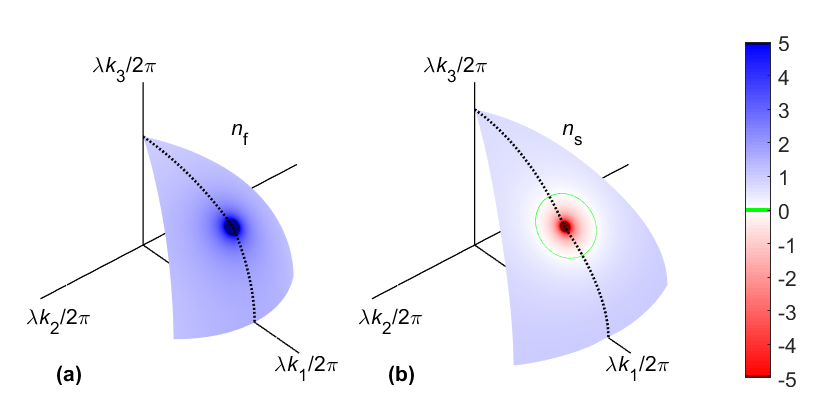}
  \caption{Color-mapped normal surface near the optic axis of a biaxial crystal: (a) fast eigenmode, $w_{\mathrm{f}1}$ as color map; (b) slow eigenmode, $w_{\mathrm{s}2}$ as color map.}\label{fig-color}
\end{figure}

The fragment of the normal surface near the optic axis of a biaxial crystal is shown in Fig.~\ref{fig-color}. Hereinafter, we use subscript ``f'' to denote quantities related to the fast eigenwave and ``s'' to denote quantities related to the slow eigenwave. One can see that far from the optic axis beam divergence of both eigenmodes is slightly anisotropic, and the diffraction coefficients are close to 1. Beam divergence is strongly anisotropic only near the optic axis. One can see that the locus of autocollimation points ($w_2$=0) is a cone around the optic axis. The diffraction coefficients become infinite approaching the optic axis because the normal surface is not differentiable in a conical degeneracy point. The directions with $w(\psi)=0$ given by~\eqref{eq-psi0} exist inside the cone $w_{\mathrm{s}2}=0$ because $w_{\mathrm{s}1}>0$.

\subsection{Acousto-optic tunable filters}\label{sec-npm}
AOTF is a type of photonic devices based on Bragg diffraction in birefringent crystals. NPM  geometries of anisotropic Bragg diffraction are commonly used to provide wide acceptance angle in AOTFs that makes possible image processing. Detailed analysis of NPM geometry in paratellurite and its applications in AOTFs was performed by Chang~\cite{Chang77}. Numerical analysis for other uniaxial acousto-optic crystals was performed by Voloshinov and Mosquera~\cite{VoloshinovMosquera06_eng}. %Specific configurations of phase-matching in noncollinear AOTFs can be used for phase imaging and laser beam shaping~\cite{YushkovChampagneKastelik20,YushkovChizhikovMakarovMolchanov20_AO}.

\begin{figure}
  \centering
  \includegraphics[width=\columnwidth]{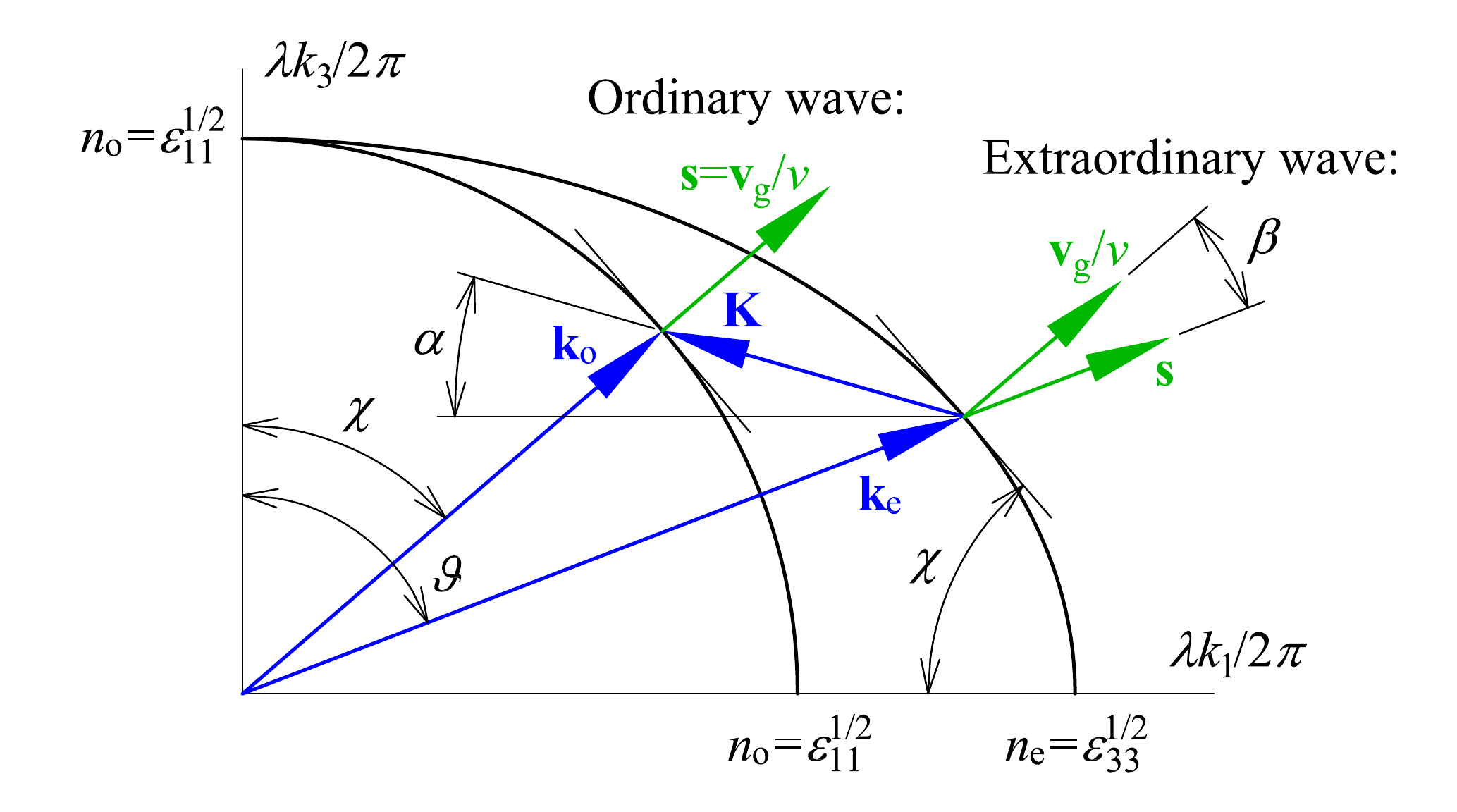}
  \caption{Wave vector diagram of wide-angle noncollinear acousto-optic diffraction in a uniaxial crystal.}\label{fig-vd}
\end{figure}

Wave vector diagram of NPM geometry in a uniaxial crystal is shown in Fig.~\ref{fig-vd}. All wave vectors are normalized to the vacuum optical wavenumber $k=2\pi/\lambda$ (where $\lambda$ is the optical wavelength). Here $k_3$ is the crystal symmetry axis [001] and $k_1$ is one of orthogonal axes. The optical normal surface section for the extraordinary wave is an ellipse given by
\begin{equation}\label{eq-ne}
  n_{\rm e}(\vartheta) = \sqrt{\frac{\varepsilon_{11}\varepsilon_{33}}{\varepsilon_{11}\sin^2\vartheta + \varepsilon_{33}\cos^2\vartheta}}.
\end{equation}
The extraordinary wave vector $\vec k_{\rm e}$ makes an angle $\vartheta$ with the optic axis of the crystal. The angle $\chi$ of ordinary wave vector $\vec k_{\rm o}$ with $k_3$ axis is found from the parallel tangent condition:
\begin{equation}\label{eq-chi}
  \chi = \arctan\left(\frac{\varepsilon_{11}}{\varepsilon_{33}}\tan\vartheta\right).
\end{equation}
The acoustic wave vector $\vec K$ is tilted by the angle
\begin{equation}\label{eq-alpha}
  \alpha = \arctan\frac{\sqrt{\varepsilon_{11}}\cos\chi - n_{\rm e}(\vartheta) \cos\vartheta}{\sqrt{\varepsilon_{11}}\sin\chi - n_{\rm e}(\vartheta) \sin\vartheta}.
\end{equation}

It has been shown that $\alpha(\vartheta)$ is a nonmonotonous function with one maximum $\alpha_{max}$ corresponding to cubic local dependence of phase-matched ultrasound frequency $F= K V/2\pi$, on optical incidence angle $\vartheta$~\cite{Chang77}, where  $K = |\vec K|$ and $V = V(\alpha)$ is the acoustic phase velocity. The point where $F(\vartheta)$ is a cubic function (i.e. $F'(\vartheta)=0$ and $F''(\vartheta)=0$) has two equivalent interpretations. The first, it is the maximum of $\alpha(\vartheta)$ given by (\ref{eq-alpha}). The second, it is a point where the curvature of the ellipsoid of the extraordinary wave normal surface equals to that of sphere of  the ordinary wave.

The ordinary wave in a uniaxial crystal has isotropic behavior. Thus, for it $\vec g=0$, and both eigenvalues of $\widehat W$ are the same, $w_1 = w_2=1$, and any curvature radius is $R_{\rm o}=\varepsilon_{11}^{1/2}$.

The curvature radii of the extraordinary wave normal surface can be analytically expressed as
\begin{equation}\label{eq-r1}
  R_{\rm p} = \frac{n_{\rm e}(\vartheta)}{w_1(\vartheta)}=  \frac{(\varepsilon_{11}^2 + (\varepsilon_{33} - \varepsilon_{11})n_{\rm e}^2(\vartheta)\cos^2\vartheta)^{3/2}}{\varepsilon_{11}^2 \varepsilon_{33}^{1/2}}
\end{equation}
 and
\begin{equation}\label{eq-r2}
  R_{\rm s} = \frac{n_{\rm e}(\vartheta)}{w_2(\vartheta)}= \frac{\varepsilon_{33}}{n_{\rm e}(\vartheta)}.
\end{equation}
where $R_{\rm p}$ is the curvature radius in the principal plane and $R_{\rm s}$ is the orthogonal (sagittal) radius.

$R_{\rm p}(\vartheta)$ is a monotonic function ranging from $R_{\rm p}(0) = \varepsilon_{33}/\varepsilon_{11}^{1/2}$ to $R_{\rm p}(\pi/2) = \varepsilon_{11}/\varepsilon_{33}^{1/2}$. Since
\begin{equation}\label{eq-comp}
  \min R_{\rm p}(\vartheta) < \varepsilon_{11}^{1/2} < \max R_{\rm p}(\vartheta),
\end{equation}
there exist one point where
\begin{equation}\label{eq-eq}
 R_{\rm p}(\vartheta) = \varepsilon_{11}^{1/2}.
\end{equation}
Equal curvature radii of the normal surface sections corresponding to $\vec k_{\rm o}$ and $\vec k_{\rm e}$ wave vectors mean that the first and the second derivatives or the surfaces are the same.

Solving Eq.~(\ref{eq-eq}) with $R$ given by (\ref{eq-r1}) yields the root
\begin{equation}\label{eq-Thetamax}
  \vartheta_{max} = \arctan\sqrt{\frac{1 + (\varepsilon_{11}/\varepsilon_{33})^{1/3}}{(\varepsilon_{11}/\varepsilon_{33})^{5/3}}}.
\end{equation}
The corresponding value of acoustic propagation angle $\alpha$ is
\begin{equation}\label{eq1-AlphaMax}
  \alpha_{max} = \arctan\sqrt{\frac{\varepsilon_{11}}{\left({\varepsilon_{11}}^{1/3}+{\varepsilon_{33}}^{1/3}\right)^3}}.
\end{equation}
One can easily verify that $\vartheta_{max}$ and $\alpha_{max}$ exactly correspond to the values obtained by explicit differentiating (\ref{eq-alpha}) and searching for the maximum as $d\alpha/d\vartheta =0 $~\cite{PozharMachihin12}.

\begin{figure}
  \centering
  \includegraphics[width=\columnwidth]{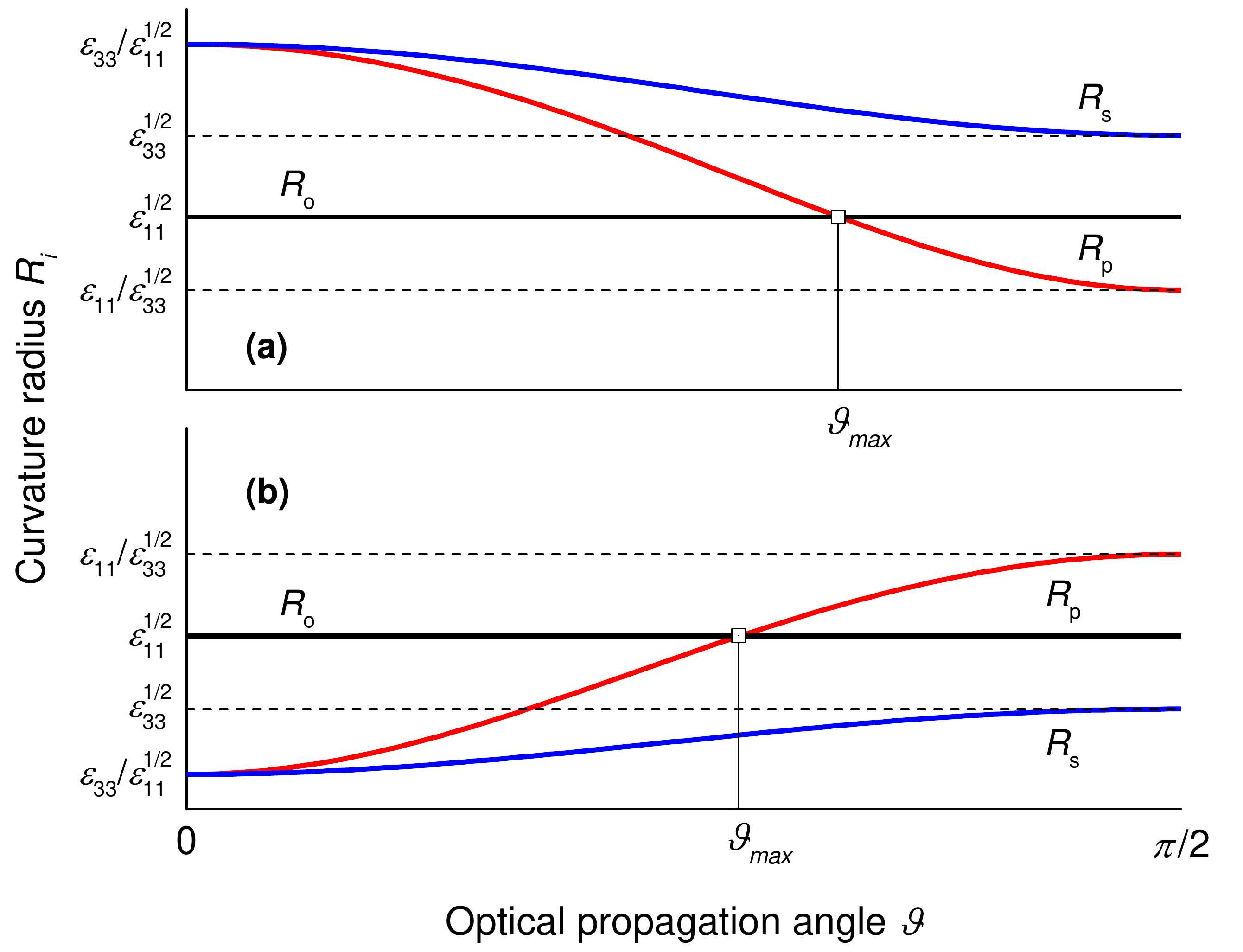}
  \caption{Curvature radii of the optical normal surface in a uniaxial crystal: (a) positive crystal $\varepsilon_{11}<\varepsilon_{33}$;  (a) negative crystal $\varepsilon_{11}>\varepsilon_{33}$.}\label{fig-R}
\end{figure}

\begin{figure}
  \centering
  \includegraphics[width=\columnwidth]{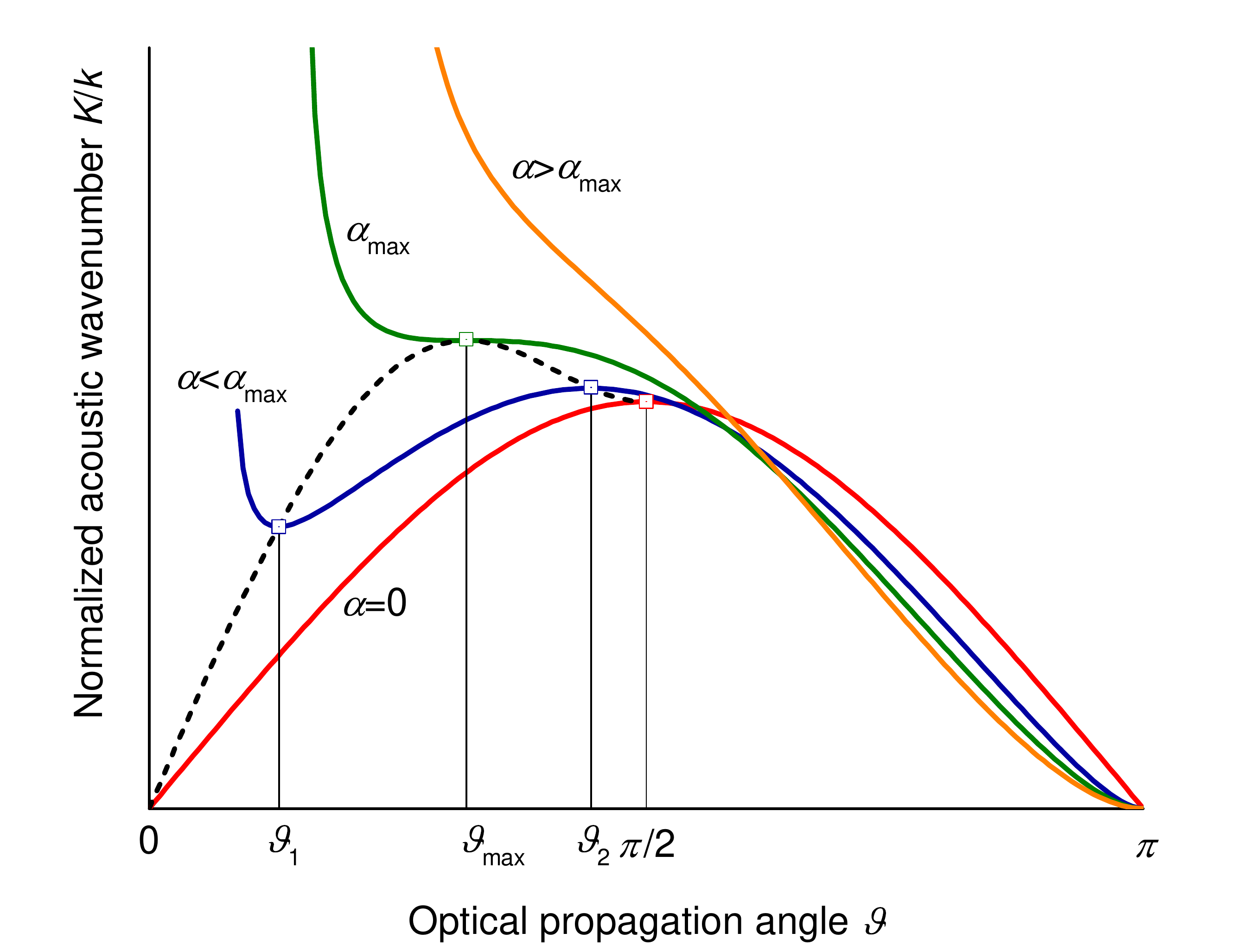}
  \caption{NPM points of acousto-optic diffraction as local minima and maxima of ultrasound frequency in a uniaxial crystal. Dashed line~--- locus of NPM points.}\label{fig-f}
\end{figure}

There are two extreme points of $F(\vartheta)$ at any $\alpha\in(0,\alpha_{max})$. One of them at $\vartheta_1\in(0,\vartheta_{max})$ is the local minimum of the frequency, and the other at $\vartheta_2\in(\vartheta_{max},\pi/2)$ is the local maximum. The points merge at $\vartheta = \vartheta_{\max}$ and $\alpha=\alpha_{max}$. Sample plots of normalized acoustic wavenumber $K(\vartheta)/k$ at different $\alpha$ are shown in Fig.~\ref{fig-f}.

Analysis of the curvature radii of the extraordinary branch of the normal surface (Fig.~\ref{fig-R}) reveals the transfer function of a noncollinear AOTF has the same widths in orthogonal directions, when $\vartheta\rightarrow0$. The width of phase matching region is proportional to $(R_i - R_{\rm o})^{-1}$, apart from the case $R_{\rm p} = R_{\rm o}$, when higher order series terms are effective. As follows from~\eqref{eq-r1} and \eqref{eq-r2} (see Fig.~\ref{fig-R}),
\begin{equation}\label{eq-lims}
  \lim_{\vartheta\rightarrow0}R_{\rm p} = \lim_{\vartheta\rightarrow0}R_{\rm s} = \varepsilon_{33}/\varepsilon_{11}^{1/2}.
\end{equation}
At first glance, this seems to contradict with the transfer function simulations made by Pozhar and Machikhin~\cite{PozharMachihin12}, who claimed that the AOTF angular aperture in the direction orthogonal to the diffraction plane sufficiently increases as $\vartheta\rightarrow0$ (or at $\theta_1\rightarrow\pi/2$ using the notation $\theta_1=\pi/2-\vartheta$ as in~\cite{PozharMachihin12}). Accurate analysis shows that the Euler angles in crystallographic axes were used by Pozhar and Machikhin as coordinates of for transfer function calculations. Thus, the azimuthal axis in those simulations should be scaled by the factor $\sin\vartheta$ to obtain the width of the transfer function correlating with the experimental results published elsewhere~\cite{BalakshyKostyuk09,YushkovMolchanovBelousovAbrosimov16,SPIE18_10744}.

\subsection{Noncritical phase matching in nonlinear optics}

Three-wave mixing is an important class of nonlinear optical interactions, which includes SHG ($\omega+\omega=2\omega$), THG ($\omega+2\omega=3\omega$), and OPA ($\omega_1+\omega_2=\omega_3$) processes. Phase matching in birefringent crystals is crucial for obtaining high coupling efficiency in applications. There are different types of NPM in three-wave mixing, including angular, spectral, temperature noncritical configurations. Angular NPM is used to provide high efficiency conversion on a wide angular aperture in. Besides that, angular NPM eliminates walk-off between interacting waves. It exists both for type-I (slow-slow-fast) and type-II (slow-fast-fast) interactions.
NPM geometry of OPA can be used for imaging applications~\cite{DevauxLantz95,VaughanTrebino11,ZengCaiChen16}. The following example demonstrates that the diffraction tensor is a simple tool for analysis of angular NPM geometries in nonlinear optics and  prediction of high-order NPM configurations.

A variety of SHG geometries in biaxial crystals can be classified using topology  of phase matching loci in stereographic projections~\cite{Hobden67,StepanovShigorinShipulo84_eng,GrechinDmitriev00_eng}. Similar diagrams are used to describe quasi-phase-matching in periodically-poled crystals~\cite{PetitBoulangerSegondsTaira07}. The method of projections helps to predict ... and find angular NPM points. In a general case, NPM points in biaxial crystals are lay out of crystal symmetry planes.

\begin{figure}
  \centering
  \includegraphics[width=\columnwidth]{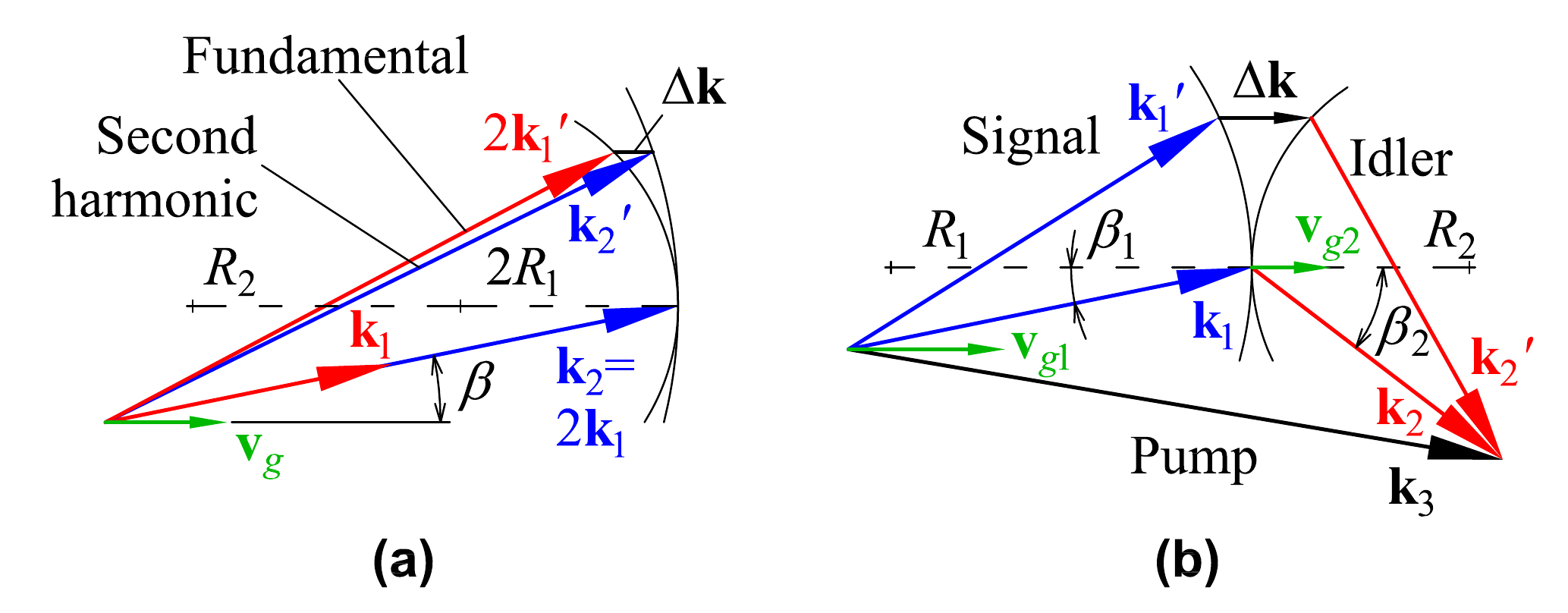}
  \caption{Wave vector diagrams in NPM geometry: (a) type-I collinear SHG; (b) noncollinear OPA.}\label{fig-npm}
\end{figure}

The wave vector diagrams of type-I collinear SHG and noncollinear OPA are shown in Fig.~\ref{fig-npm}. In both cases, the phase mismatch vector $\Delta\vec k$ is parallel to the common direction of group velocity vectors. The magnitude of mismatch $\Delta k=|\Delta\vec k|$ depends is a quadratic function of the angular deviation from phase matching.

The mismatch vector in the SHG case, Fig.~\ref{fig-npm} (a), is
\begin{equation}\label{eq-mm-shg}
  \Delta\vec k  = \vec k_{2}' - 2\vec k_{1}',
\end{equation}
where subscripts 1 and 2  denote the fundamental and the second harmonic waves. The walk-off angle $\beta$ is the same for them, but the curvature radii of the slowness surfaces are different.
The phase mismatch magnitude is $\Delta k\propto |R_2 - 2R_1|$.

The mismatch vector in the OPA case, Fig.~\ref{fig-npm} (b), is
\begin{equation}\label{eq-mm-nopa}
  \Delta\vec k  = \vec k_3 - (\vec k_1' + \vec k_2'),
\end{equation}
where subscripts 1 and 2 denote the signal and the idler, respectively, and 3 denotes the pump, which is assumed to be collimated. The  phase mismatch magnitude is  $\Delta k\propto |R_1+R_2|$.

\begin{figure}[t]
  \centering
  \includegraphics[width=\columnwidth]{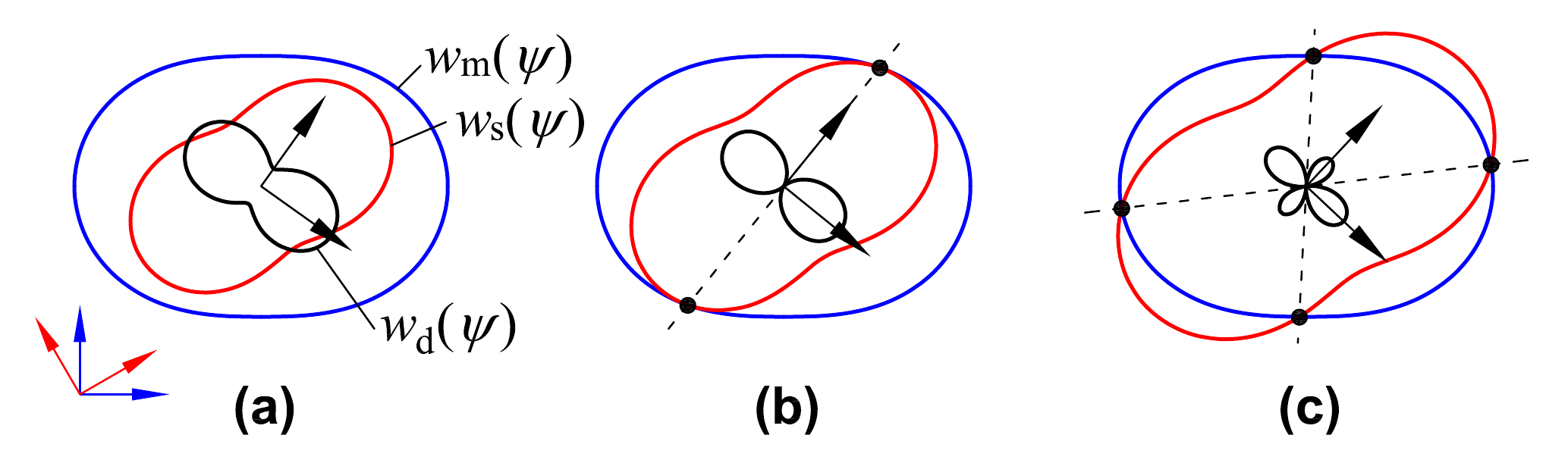}
  \caption{Difference of two curvature tensors is also a planar tensor, and its zero directions correspond to high-order NPM configurations: (a) no axes of high-order NPM; (b) one axis of high-order NPM; (c) two axes of high-order NPM. }\label{fig-Ris}
\end{figure}

Since any linear combination of planar tensors is also a planar tensor, its zero directions can be used to find higher-order NPM configurations. This case is illustrated for a difference of two tensors in Fig.~\ref{fig-Ris}. The difference tensor $\widehat W_{\rm d}$ is
\begin{equation}\label{eq-Wd}
  \widehat W_{\rm d} = \widehat W_{\rm m} - \widehat W_{\rm s},
\end{equation}
where $\widehat W_{\rm m}$ and $\widehat W_{\rm s}$ are arbitrary planar second rank tensor (the plot shows the case $w_{{\rm m}1}w_{{\rm m}2}>0$ and $w_{{\rm s}1}w_{{\rm s}2}>0$, but a general case will give the same results). Three possible cases are for $\widehat W_{\rm d}$:

(a) $w_{\rm m}(\psi)>w_{\rm s}(\psi)$ or $w_{\rm m}(\psi)<w_{\rm s}(\psi)$ for all $\psi$, $w_{{\rm d}1}w_{{\rm d}2}>0$;

(b) $w_{\rm m}(\psi)=w_{\rm s}(\psi)$ at a single value of $\psi$ corresponding to the eigenvector of $\widehat W_{\rm d}$, $w_{{\rm d}1}w_{{\rm d}2}=0$;

(c) $w_{\rm m}(\psi)=w_{\rm s}(\psi)$ at two values of $\psi$, $w_{{\rm d}1}w_{{\rm d}2}<0$.

The case (a) is a simple NPM configuration, and the phase mismatch $\Delta k$ is a quadratic function of angular deviation for any azimuthal angle $\psi$. The transfer function topology in this case is O-type. The case (b) corresponds to one direction of high-order NPM geometry along the eigenvector of $\widehat W_{\rm d}$ with zero eigenvalue. The transfer function topology in this case is I-type. The case (c) corresponds to two crossed directions of high-order NPM geometry with directions given by \eqref{eq-psi0} for the tensor $\widehat W_{\rm d}$. The transfer function topology in this case is X-type. Note that the same three topologies of the transfer function do exist also for anisotropic acousto-optic interaction, which is described by a similar wave vector diagram approach as nonlinear three-wave mixing~\cite{SPIE18_10744}.

\section{Conclusion}

In this paper, we defined the beam diffraction tensor using the transverse component of the group velocity vector and its derivatives. The eigenvectors of the tensor determine the transverse directions of maximum and minimum far field beam divergence. The eigenvalues of the tensor determine scaling of the Green's function, which in turn affects the beam divergence and geometry of the wavefronts.

\section*{Acknowledgments}

The research was supported by the Russian Foundation for Basic Research (Project 21-12-00247).


\begin{thebibliography}{32}
\expandafter\ifx\csname natexlab\endcsname\relax\def\natexlab#1{#1}\fi
\expandafter\ifx\csname bibnamefont\endcsname\relax
  \def\bibnamefont#1{#1}\fi
\expandafter\ifx\csname bibfnamefont\endcsname\relax
  \def\bibfnamefont#1{#1}\fi
\expandafter\ifx\csname citenamefont\endcsname\relax
  \def\citenamefont#1{#1}\fi
\expandafter\ifx\csname url\endcsname\relax
  \def\url#1{\texttt{#1}}\fi
\expandafter\ifx\csname urlprefix\endcsname\relax\def\urlprefix{URL }\fi
\providecommand{\bibinfo}[2]{#2}
\providecommand{\eprint}[2][]{\url{#2}}

\bibitem[{\citenamefont{{A. Yariv} and {P. Yeh}}(1984)}]{YarivYeh}
\bibinfo{author}{\bibnamefont{{A. Yariv}}} \bibnamefont{and}
  \bibinfo{author}{\bibnamefont{{P. Yeh}}}, \emph{\bibinfo{title}{{Optical
  Waves in Crystals}}} (\bibinfo{publisher}{{Wiley}}, \bibinfo{address}{New
  York}, \bibinfo{year}{1984}).

\bibitem[{\citenamefont{{Q. Rolland} et~al.}(2014)\citenamefont{{Q. Rolland},
  {S. Dupont}, {J. Gazalet}, {J.-C. Kastelik}, {Y. Pennec}, {B.
  Djafari-Rouhani}, and {V. Laude}}}]{RollandDupontGazalet14}
\bibinfo{author}{\bibnamefont{{Q. Rolland}}}, \bibinfo{author}{\bibnamefont{{S.
  Dupont}}}, \bibinfo{author}{\bibnamefont{{J. Gazalet}}},
  \bibinfo{author}{\bibnamefont{{J.-C. Kastelik}}},
  \bibinfo{author}{\bibnamefont{{Y. Pennec}}},
  \bibinfo{author}{\bibnamefont{{B. Djafari-Rouhani}}}, \bibnamefont{and}
  \bibinfo{author}{\bibnamefont{{V. Laude}}}, \bibinfo{journal}{Opt. Express}
  \textbf{\bibinfo{volume}{22}}, \bibinfo{pages}{16288} (\bibinfo{year}{2014}).

\bibitem[{\citenamefont{{Y. Petit} et~al.}(2013)\citenamefont{{Y. Petit}, {S.
  Joly}, {P. Segonds}, and {B. Boulanger}}}]{PetitJolySegondsBoulanger13}
\bibinfo{author}{\bibnamefont{{Y. Petit}}}, \bibinfo{author}{\bibnamefont{{S.
  Joly}}}, \bibinfo{author}{\bibnamefont{{P. Segonds}}}, \bibnamefont{and}
  \bibinfo{author}{\bibnamefont{{B. Boulanger}}}, \bibinfo{journal}{Laser
  Photonics Rev.} \textbf{\bibinfo{volume}{7}}, \bibinfo{pages}{920}
  (\bibinfo{year}{2013}).

\bibitem[{\citenamefont{{A. Turpin} et~al.}(2016)\citenamefont{{A. Turpin},
  {Y.V. Loiko}, {T.K. Kalkandjiev}, and {J.
  Mompart}}}]{TurpinLoikoKalkandjievMompart16}
\bibinfo{author}{\bibnamefont{{A. Turpin}}},
  \bibinfo{author}{\bibnamefont{{Y.V. Loiko}}},
  \bibinfo{author}{\bibnamefont{{T.K. Kalkandjiev}}}, \bibnamefont{and}
  \bibinfo{author}{\bibnamefont{{J. Mompart}}}, \bibinfo{journal}{Laser
  Photonics Rev.} \textbf{\bibinfo{volume}{10}}, \bibinfo{pages}{750}
  (\bibinfo{year}{2016}).

\bibitem[{\citenamefont{{V.N. Belyi} et~al.}(2016)\citenamefont{{V.N. Belyi},
  {P.A. Khilo}, {N.S. Kazak}, and {N.A. Khilo}}}]{BelyiKhiloKazak16}
\bibinfo{author}{\bibnamefont{{V.N. Belyi}}},
  \bibinfo{author}{\bibnamefont{{P.A. Khilo}}},
  \bibinfo{author}{\bibnamefont{{N.S. Kazak}}}, \bibnamefont{and}
  \bibinfo{author}{\bibnamefont{{N.A. Khilo}}}, \bibinfo{journal}{J. Opt.}
  \textbf{\bibinfo{volume}{18}}, \bibinfo{pages}{074002}
  (\bibinfo{year}{2016}).

\bibitem[{\citenamefont{{L. Bergstein} and {T.
  Zachos}}(1999)}]{BergsteinZachos66}
\bibinfo{author}{\bibnamefont{{L. Bergstein}}} \bibnamefont{and}
  \bibinfo{author}{\bibnamefont{{T. Zachos}}}, \bibinfo{journal}{J. Opt. Soc.
  Am.} \textbf{\bibinfo{volume}{56}}, \bibinfo{pages}{931}
  (\bibinfo{year}{1999}).

\bibitem[{\citenamefont{{N.R. Ogg}}(1971)}]{Ogg71}
\bibinfo{author}{\bibnamefont{{N.R. Ogg}}}, \bibinfo{journal}{{J. Phys. A: Gen.
  Phys.}} \textbf{\bibinfo{volume}{4}}, \bibinfo{pages}{382}
  (\bibinfo{year}{1971}).

\bibitem[{\citenamefont{{J.A. Fleck} and {M.D. Feit}}(1983)}]{FleckFeit83}
\bibinfo{author}{\bibnamefont{{J.A. Fleck}}} \bibnamefont{and}
  \bibinfo{author}{\bibnamefont{{M.D. Feit}}}, \bibinfo{journal}{J. Opt. Soc.
  Am.} \textbf{\bibinfo{volume}{73}}, \bibinfo{pages}{920}
  (\bibinfo{year}{1983}).

\bibitem[{\citenamefont{{A. Ciattoni} et~al.}(2001)\citenamefont{{A. Ciattoni},
  {B. Crosignani}, and {P. Di Porto}}}]{CiattoniCrosignaniPorto01}
\bibinfo{author}{\bibnamefont{{A. Ciattoni}}},
  \bibinfo{author}{\bibnamefont{{B. Crosignani}}}, \bibnamefont{and}
  \bibinfo{author}{\bibnamefont{{P. Di Porto}}}, \bibinfo{journal}{J. Opt. Soc.
  Am. A~-- Opt. Image Sci. Vis.} \textbf{\bibinfo{volume}{18}},
  \bibinfo{pages}{1656} (\bibinfo{year}{2001}).

\bibitem[{\citenamefont{{M. Nawareg}}(2019)}]{Nawareg19}
\bibinfo{author}{\bibnamefont{{M. Nawareg}}}, \bibinfo{journal}{J. Opt. Soc.
  Am. B~-- Opt. Phys.} \textbf{\bibinfo{volume}{36}}, \bibinfo{pages}{470}
  (\bibinfo{year}{2019}).

\bibitem[{\citenamefont{{A.G. Khatkevich}}(1978)}]{Khatkevich78_eng}
\bibinfo{author}{\bibnamefont{{A.G. Khatkevich}}}, \bibinfo{journal}{Sov. Phys.
  Acoust.} \textbf{\bibinfo{volume}{24}}, \bibinfo{pages}{108}
  (\bibinfo{year}{1978}).

\bibitem[{\citenamefont{{N.F. Naumenko} et~al.}(1983)\citenamefont{{N.F.
  Naumenko}, {N.V. Perelomova}, and {V.S.
  Bondarenko}}}]{NaumenkoPerelomovaBondarenko83_eng}
\bibinfo{author}{\bibnamefont{{N.F. Naumenko}}},
  \bibinfo{author}{\bibnamefont{{N.V. Perelomova}}}, \bibnamefont{and}
  \bibinfo{author}{\bibnamefont{{V.S. Bondarenko}}}, \bibinfo{journal}{Sov.
  Phys. Crystallography} \textbf{\bibinfo{volume}{28}}, \bibinfo{pages}{607}
  (\bibinfo{year}{1983}).

\bibitem[{\citenamefont{{N.F. Naumenko} et~al.}(2021)\citenamefont{{N.F.
  Naumenko}, {K.B. Yushkov}, and {V.Ya.
  Molchanov}}}]{NaumenkoYushkovMolchanov21}
\bibinfo{author}{\bibnamefont{{N.F. Naumenko}}},
  \bibinfo{author}{\bibnamefont{{K.B. Yushkov}}}, \bibnamefont{and}
  \bibinfo{author}{\bibnamefont{{V.Ya. Molchanov}}}, \bibinfo{journal}{Eur.
  Phys. J. Plus} \textbf{\bibinfo{volume}{136}}, \bibinfo{pages}{95}
  (\bibinfo{year}{2021}).

\bibitem[{\citenamefont{{V.I. Alshits} and {V.N.
  Lyubimov}}(2013)}]{AlshitsLyubimov13_eng}
\bibinfo{author}{\bibnamefont{{V.I. Alshits}}} \bibnamefont{and}
  \bibinfo{author}{\bibnamefont{{V.N. Lyubimov}}}, \bibinfo{journal}{Phys.
  Usp.} \textbf{\bibinfo{volume}{56}}, \bibinfo{pages}{1021}
  (\bibinfo{year}{2013}).

\bibitem[{\citenamefont{{M.P. Do Carmo}}(2016)}]{Carmo}
\bibinfo{author}{\bibnamefont{{M.P. Do Carmo}}},
  \emph{\bibinfo{title}{{Differential Geometry of Curves and Surfaces}}}
  (\bibinfo{publisher}{{Dover Publ. Inc.}}, \bibinfo{address}{Mineola, NY},
  \bibinfo{year}{2016}), \bibinfo{edition}{2nd} ed., ISBN
  \bibinfo{isbn}{9780486806990}.

\bibitem[{\citenamefont{{A.L. Shuvalov}}(1998)}]{Shuvalov98}
\bibinfo{author}{\bibnamefont{{A.L. Shuvalov}}}, \bibinfo{journal}{Proc. Roy.
  Soc. London A} \textbf{\bibinfo{volume}{454}}, \bibinfo{pages}{2911}
  (\bibinfo{year}{1998}).

\bibitem[{\citenamefont{{N. Naumenko} et~al.}(2013)\citenamefont{{N. Naumenko},
  {S.I. Chizhikov}, {V.Ya. Molchanov}, and {K.B. Yushkov}}}]{IUS13}
\bibinfo{author}{\bibnamefont{{N. Naumenko}}},
  \bibinfo{author}{\bibnamefont{{S.I. Chizhikov}}},
  \bibinfo{author}{\bibnamefont{{V.Ya. Molchanov}}}, \bibnamefont{and}
  \bibinfo{author}{\bibnamefont{{K.B. Yushkov}}}, in
  \emph{\bibinfo{booktitle}{2013 Joint UFFC, EFTF and PFM Symposium, 2013 IUS
  Proceedings}} (\bibinfo{publisher}{IEEE}, \bibinfo{address}{Prague},
  \bibinfo{year}{2013}), pp. \bibinfo{pages}{500--503}, ISBN
  \bibinfo{isbn}{978-1-4673-5686-2}.

\bibitem[{\citenamefont{{M. Born} and {E. Wolf}}(1999)}]{BornWolf7ed}
\bibinfo{author}{\bibnamefont{{M. Born}}} \bibnamefont{and}
  \bibinfo{author}{\bibnamefont{{E. Wolf}}}, \emph{\bibinfo{title}{{Principles
  of Optics: Electromagnetic Theory of Propagation, Interference and
  Diffraction of Light}}} (\bibinfo{publisher}{{Cambridge University Press}},
  \bibinfo{address}{Cambridge}, \bibinfo{year}{1999}), \bibinfo{edition}{{7th
  (expanded)}} ed., ISBN \bibinfo{isbn}{0521642221}.

\bibitem[{\citenamefont{{J.W. Goodman}}(2005)}]{GoodmanIFO}
\bibinfo{author}{\bibnamefont{{J.W. Goodman}}},
  \emph{\bibinfo{title}{{Introduction to Fourier Optics}}}
  (\bibinfo{publisher}{{Roberts}}, \bibinfo{address}{{New York}},
  \bibinfo{year}{2005}), \bibinfo{edition}{{3rd}} ed.

\bibitem[{\citenamefont{{I.C. Chang}}(1977)}]{Chang77}
\bibinfo{author}{\bibnamefont{{I.C. Chang}}}, \bibinfo{journal}{Opt. Eng.}
  \textbf{\bibinfo{volume}{16}}, \bibinfo{pages}{455} (\bibinfo{year}{1977}).

\bibitem[{\citenamefont{{V.B. Voloshinov} and {J.C.
  Mosquera}}(2006)}]{VoloshinovMosquera06_eng}
\bibinfo{author}{\bibnamefont{{V.B. Voloshinov}}} \bibnamefont{and}
  \bibinfo{author}{\bibnamefont{{J.C. Mosquera}}}, \bibinfo{journal}{Opt.
  Spectrosc.} \textbf{\bibinfo{volume}{101}}, \bibinfo{pages}{635}
  (\bibinfo{year}{2006}).

\bibitem[{\citenamefont{{V. Pozhar} and {A.
  Machihin}}(2012)}]{PozharMachihin12}
\bibinfo{author}{\bibnamefont{{V. Pozhar}}} \bibnamefont{and}
  \bibinfo{author}{\bibnamefont{{A. Machihin}}}, \bibinfo{journal}{Appl. Opt.}
  \textbf{\bibinfo{volume}{51}}, \bibinfo{pages}{4513} (\bibinfo{year}{2012}).

\bibitem[{\citenamefont{{V.I. Balakshy} and {D.E.
  Kostyuk}}(2009)}]{BalakshyKostyuk09}
\bibinfo{author}{\bibnamefont{{V.I. Balakshy}}} \bibnamefont{and}
  \bibinfo{author}{\bibnamefont{{D.E. Kostyuk}}}, \bibinfo{journal}{Appl. Opt.}
  \textbf{\bibinfo{volume}{48}}, \bibinfo{pages}{C24} (\bibinfo{year}{2009}).

\bibitem[{\citenamefont{{K.B. Yushkov} et~al.}(2016)\citenamefont{{K.B.
  Yushkov}, {V.Ya. Molchanov}, {P.V. Belousov}, and {A.Yu.
  Abrosimov}}}]{YushkovMolchanovBelousovAbrosimov16}
\bibinfo{author}{\bibnamefont{{K.B. Yushkov}}},
  \bibinfo{author}{\bibnamefont{{V.Ya. Molchanov}}},
  \bibinfo{author}{\bibnamefont{{P.V. Belousov}}}, \bibnamefont{and}
  \bibinfo{author}{\bibnamefont{{A.Yu. Abrosimov}}}, \bibinfo{journal}{J.
  Biomed. Opt.} \textbf{\bibinfo{volume}{21}}, \bibinfo{pages}{016003}
  (\bibinfo{year}{2016}).

\bibitem[{\citenamefont{{K.B. Yushkov} et~al.}(2018)\citenamefont{{K.B.
  Yushkov}, {V.Ya. Molchanov}, {V.I. Balakshy}, and {S.N.
  Mantsevich}}}]{SPIE18_10744}
\bibinfo{author}{\bibnamefont{{K.B. Yushkov}}},
  \bibinfo{author}{\bibnamefont{{V.Ya. Molchanov}}},
  \bibinfo{author}{\bibnamefont{{V.I. Balakshy}}}, \bibnamefont{and}
  \bibinfo{author}{\bibnamefont{{S.N. Mantsevich}}}, in
  \emph{\bibinfo{booktitle}{Laser Beam Shaping XVIII}}, edited by
  \bibinfo{editor}{\bibnamefont{{A. Dudley}}} \bibnamefont{and}
  \bibinfo{editor}{\bibnamefont{{A.V. Laskin}}} (\bibinfo{publisher}{SPIE},
  \bibinfo{year}{2018}), vol. \bibinfo{volume}{10744} of
  \emph{\bibinfo{series}{Proc. SPIE}}, p. \bibinfo{pages}{107440Q}.

\bibitem[{\citenamefont{{F. Devaux} and {E. Lantz}}(1995)}]{DevauxLantz95}
\bibinfo{author}{\bibnamefont{{F. Devaux}}} \bibnamefont{and}
  \bibinfo{author}{\bibnamefont{{E. Lantz}}}, \bibinfo{journal}{J. Opt. Soc.
  Am. B~-- Opt. Phys.} \textbf{\bibinfo{volume}{12}}, \bibinfo{pages}{2245}
  (\bibinfo{year}{1995}).

\bibitem[{\citenamefont{{J.C. Vaughan} and {R.
  Trebino}}(2011)}]{VaughanTrebino11}
\bibinfo{author}{\bibnamefont{{J.C. Vaughan}}} \bibnamefont{and}
  \bibinfo{author}{\bibnamefont{{R. Trebino}}}, \bibinfo{journal}{Opt. Express}
  \textbf{\bibinfo{volume}{19}}, \bibinfo{pages}{8920} (\bibinfo{year}{2011}).

\bibitem[{\citenamefont{{X. Zeng} et~al.}({2016})\citenamefont{{X. Zeng}, {Y.
  Cai}, {W. Chen}, {J. Li}, {S. Zheng}, {T. Zhu}, and {S. Xu}}}]{ZengCaiChen16}
\bibinfo{author}{\bibnamefont{{X. Zeng}}}, \bibinfo{author}{\bibnamefont{{Y.
  Cai}}}, \bibinfo{author}{\bibnamefont{{W. Chen}}},
  \bibinfo{author}{\bibnamefont{{J. Li}}}, \bibinfo{author}{\bibnamefont{{S.
  Zheng}}}, \bibinfo{author}{\bibnamefont{{T. Zhu}}}, \bibnamefont{and}
  \bibinfo{author}{\bibnamefont{{S. Xu}}}, \bibinfo{journal}{IEEE Photonics
  Technol. Lett.} \textbf{\bibinfo{volume}{28}}, \bibinfo{pages}{2685}
  (\bibinfo{year}{{2016}}).

\bibitem[{\citenamefont{{M.V. Hobden}}(1967)}]{Hobden67}
\bibinfo{author}{\bibnamefont{{M.V. Hobden}}}, \bibinfo{journal}{J. Appl.
  Phys.} \textbf{\bibinfo{volume}{38}}, \bibinfo{pages}{4365}
  (\bibinfo{year}{1967}).

\bibitem[{\citenamefont{{D.Yu. Stepanov} et~al.}(1984)\citenamefont{{D.Yu.
  Stepanov}, {V.D. Shigorin}, and {G.P.
  Shipulo}}}]{StepanovShigorinShipulo84_eng}
\bibinfo{author}{\bibnamefont{{D.Yu. Stepanov}}},
  \bibinfo{author}{\bibnamefont{{V.D. Shigorin}}}, \bibnamefont{and}
  \bibinfo{author}{\bibnamefont{{G.P. Shipulo}}}, \bibinfo{journal}{Sov. J.
  Quantum Electron.} \textbf{\bibinfo{volume}{10}}, \bibinfo{pages}{1315}
  (\bibinfo{year}{1984}).

\bibitem[{\citenamefont{{S.G. Grechin} et~al.}(2000)\citenamefont{{S.G.
  Grechin}, {S.S. Grechin}, and {V.G. Dmitriev}}}]{GrechinDmitriev00_eng}
\bibinfo{author}{\bibnamefont{{S.G. Grechin}}},
  \bibinfo{author}{\bibnamefont{{S.S. Grechin}}}, \bibnamefont{and}
  \bibinfo{author}{\bibnamefont{{V.G. Dmitriev}}}, \bibinfo{journal}{Quantum
  Electron.} \textbf{\bibinfo{volume}{30}}, \bibinfo{pages}{377}
  (\bibinfo{year}{2000}).

\bibitem[{\citenamefont{{Y. Petit} et~al.}(2007)\citenamefont{{Y. Petit}, {B.
  Boulanger}, {P. Segonds}, and {T. Taira}}}]{PetitBoulangerSegondsTaira07}
\bibinfo{author}{\bibnamefont{{Y. Petit}}}, \bibinfo{author}{\bibnamefont{{B.
  Boulanger}}}, \bibinfo{author}{\bibnamefont{{P. Segonds}}}, \bibnamefont{and}
  \bibinfo{author}{\bibnamefont{{T. Taira}}}, \bibinfo{journal}{Phys. Rev. A}
  \textbf{\bibinfo{volume}{76}}, \bibinfo{pages}{063817}
  (\bibinfo{year}{2007}).

\end{thebibliography}
\end{document}